\documentclass[twocolumn,showpacs,preprintnumbers,amsmath,amssymb]{revtex4}

\usepackage{graphicx}
\usepackage{dcolumn}
\usepackage{bm}
\usepackage{color}
\newcommand{\comment}[1]{}
\def\COMMENTS{
\renewcommand{\comment}[1]{\\\textcolor{red}{\emph{##1}}\\}}
\COMMENTS

\newcommand{\Eq}{Eq.}
\newcommand{\Eqs}{Eqs.}
\newcommand{\Fig}{Fig.}

\newcommand{\ignore}[1]{}

\begin{document}


\title{Multi-component gap solitons in spinor Bose-Einstein condensates}

\author{Beata J. D\c{a}browska-W\"uster}

\author{Elena A. Ostrovskaya}

\author{Tristram J. Alexander}

\author{Yuri S. Kivshar}
\affiliation{ARC Centre of Excellence for Quantum-Atom Optics}
\affiliation{Nonlinear Physics Centre, Research School of Physical Sciences and Engineering, The Australian National University, Canberra ACT 0200, Australia}

\date{\today}

\begin{abstract}
We model the nonlinear behaviour of spin-1 Bose-Einstein condensates (BECs) with repulsive spin-independent interactions 
and either {\em ferromagnetic} or {\em anti-ferromagnetic} (polar) spin-dependent interactions, loaded into a one-dimensional optical lattice potential. We show that both types of BECs exhibit dynamical instabilities and may form spatially localized multi-component structures.
The localized states of the spinor matter waves take the form of \emph{vector gap solitons} and \emph{self-trapped waves} that exist only within gaps of the linear Bloch-wave band-gap spectrum. Of special interest  are the nonlinear localized states that do not exhibit a common spatial density profile shared by all condensate components, and consequently cannot be described by the {\it single mode approximation} (SMA) frequently employed within the framework of the mean-field treatment. We show that the non-SMA states can exhibits Josephson-like internal oscillations and {\em self-magnetisation}, i.e. intrinsic precession of the local spin. Finally, we demonstrate that non-stationary states of a spinor BEC in a lattice exhibit coherent undamped spin-mixing dynamics, and that their controlled conversion into a stationary state can be achieved by the application of an external magnetic field.
\end{abstract}

\pacs{03.75.Lm,03.75.Mn}
\maketitle

\section{\label{sec:intro}Introduction}
The development of optical trapping techniques for Bose-Einstein condensates (BECs) has enabled the confinement of atoms independently of their spin orientation in so-called \emph{spinor condensates}~\cite{stamper,chapman}. Diverse experimental and theoretical activities had aimed to characterise the effect of the spin degree of freedom on the macroscopic quantum state. A multitude of exciting phenomena which are not present in a single component BEC have been revealed. These include spin textures~\cite{vortex_exp,vortex_th}, spin domain formation~\cite{strenger,domain_th} and multi-component solitons~\cite{wadati1,wadati2,malomed_spinor}. One of the most frequently used optical traps for ultra-cold matter is an optical lattice, i.e.~a~standing light wave far detuned from the atomic resonance. 
However, up to now, the effect of the spin degree of freedom on the key properties of BECs in periodic potentials has not been fully explored.

\begin{figure}[hb]
\includegraphics[width=10\textwidth,height=5.5cm,keepaspectratio]{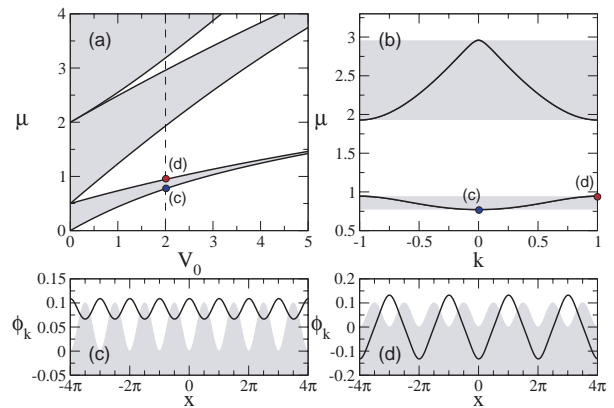}
\caption{\label{fig:bgd}
Band-gap structure of the linear Bloch-wave spectrum for a periodic potential of a one-dimensional optical lattice $V_{\rm OL}=V_0\sin^2(x)$.(a) Variation of bands (shaded) and gaps (clear areas) with the potential depth, $\mu(V_0)$, and (b) eigenvalue spectrum $\mu(k)$ reduced to the first Brillouin zone shown for $V_0=2$. Points (c) and (d) correspond to the ground and excited state linear Bloch waves with the typical spatial density profiles in a single hyperfine component of the condensate shown in panels (c) and (d), respectively.}
\end{figure}

A unique property arising due to the interplay of the inherent nonlinearity of matter waves and the periodicity of the optical lattice is the spatial localization of a BEC with {\em repulsive} inter-atomic interactions. Bragg scattering of matter waves with repulsive atomic interactions can reverse the effective diffraction properties 
of the BEC wavepackets and lead to formation of localized states with energy eigenvalues within the gaps of the linear Bloch-wave spectrum of the lattice [see \Fig~\ref{fig:bgd}(a,b)]. Two types of such localized structures have been studied in the recent years - \emph{gap solitons} and {\em self-trapped states}. Single component gap solitons in a one-dimensional (1D) lattice have been extensively studied theoretically~\cite{zobay,pearl,efrem,konotop,sanpera,generation_th} and observed experimentally~\cite{gap_soliton}. Two-component gap solitons in lattices have also been investigated \cite{elena_prl,malomed}. These nonlinearly localized states are restricted to low particle numbers. In contrast, the experimentally observed self-trapped states \cite{st_exp} can confine large number of condensate atoms \cite{st_exp,selftrapped,st_theory}, and can be interpreted as {\em gap waves} - truncated Bloch states that link gap solitons with spatially extended nonlinear Bloch waves in the lattice \cite{selftrapped}.

In this paper we present a detailed analysis of the nonlinear localization of spinor (three-component) BECs in the presence of a 1D lattice, within the framework of the mean-field theory. We consider BEC clouds with either a {\em ferromagnetic} ($^{87}{\rm Rb}$) or {\em polar} ($^{23}{\rm Na}$) spin-dependent interaction, and show that the lattice enables the existence of multi-component localized states of both types of the spinor condensates. We show that the spinor states in both of the two condensate types may be localized as \emph{vector gap solitons} that can exhibit either ferromagnetic or polar spin structures regardless of the type of spin-dependent interaction. In addition, we find and characterise the multi-component self-trapped gap states that naturally extend the concept of single-component self-trapped gap waves \cite{selftrapped}.

Both the spinor gap solitons and gap waves can be adequately described within the {\em single-mode approximation} (SMA) which assumes that all components of the composite localized state share the same spatial profile. However, we also reveal that both gap solitons and gap wave states may exist simultaneously in different spin components forming a hybrid localized state, which no longer satisfies the SMA. Such composite states may exhibit effects of self-magnetisation and internal Josephson-like dynamics \cite{Salerno_J}. 

Furthermore, we consider non-stationary dynamics of the spinor gap solitons and examine two distinct methods to induce coherent spin-mixing oscillations, through a change in the relative phase of the condensate, or through a population imbalance in a stationary state configuration. We find that the presence of the lattice potential prevents fragmentation of a spinor BEC and damping of the coherent spin-mixing evolution. Finally, we demonstrate that application of a magnetic field \cite{chapman_nature} can arrest the spin mixing, which lays the foundations for controlled and flexible state manipulation.

The paper is organized as follows: section~\ref{sec:model}, where we introduce our model is followed by section~\ref{sec:states} in which basic properties of stationary states are analysed. In section~\ref{sec:bloch} we examine the dynamical instability of spinor Bloch waves and spinor matter wave localization. Sections~\ref{sec:solitons}, \ref{sec:sts} and \ref{sec:nonsma} concern distinct types of localized stationary states. And finally, in section~\ref{sec:dynamics} coherent spin-mixing dynamics in multi-component localized states is discussed.

\section{\label{sec:model}Model}
In the framework of mean-field theory, the wavefunctions, $\Psi_{\pm 1,0} \equiv \Psi({\bf r})_{\pm 1,0}$, of the three hyperfine components ($m_{F}=\pm 1, 0$) of an  $F=1$ spinor condensate obey the following set of coupled Gross-Pitaevskii equations~\cite{marti}:
\begin{align}
\label{eq:TDGPEs}
&         i \frac {\partial \Psi_{\pm 1}}{\partial t} =
         \mathcal{L}  \Psi_{\pm1}+ \lambda_a \left( |\Psi_{\pm 1}|^2 + |\Psi_{0}|^2 - |\Psi_{\mp 1}|^2 \right) \Psi_{\pm 1}
\nonumber
\\
&
\;\;\;\;\;\;\;\;\;\;\;\;\;\;\;\; + \lambda_a \Psi_{0}^2\Psi_{\mp 1}^{*}, 
\nonumber
\\
&         i \frac {\partial \Psi_{0}}{\partial t}=
         \mathcal{L} \Psi_{0} 
+ \lambda_a \left( |\Psi_{-1}|^2 + |\Psi_{+1}|^2 \right) \Psi_{0}
\\
&
\;\;\;\;\;\;\;\;\;\;\;\;\;\;\;\; + 2 \lambda_a \Psi_{+1}\Psi_{-1}\Psi_{0}^{*},
\nonumber
\end{align}
where $\mathcal{L}=-\frac{1}{2}\nabla^2 + V({\bf r}) + \lambda_s \left( |\Psi_{-1}|^{2} + |\Psi_{0}|^{2} + |\Psi_{+ 1}|^{2} \right)$. We assume the confining potential potential to be of the form $V({\bf r})=\frac{1}{2}\omega_{\perp}^{2}(y^{2}+z^{2})+V_{\rm OL}(x)$, where $\omega_{\perp}$ is the trapping frequency in the direction perpendicular to the lattice, and $V_{\rm OL}=V_{0} \sin^2(\pi x/d)$ is a periodic potential created by a 1D optical lattice, with the period ~$d$. The wavefunctions are normalised to the number of atoms: $\int d{\bf r} \:|\Psi_j|^2=N_j$, where $j=\pm1,0$ and $\Sigma N_j=\mathcal{N}$ is the total number of atoms in a condensate. Equations~(\ref{eq:TDGPEs}) were made dimensionless by measuring length, time and energy in units: $a_\perp\equiv(\hbar/m \omega_{\perp})^{\frac{1}{2}}$, $\omega_{\perp}^{-1}$, and $\hbar\omega_{\perp}$, respectively. In this work we consider both condensates of $^{87}\mbox{Rb}$ and $^{23}\mbox{Na}$ atoms, and assume strongly anisotropic clouds with transverse trapping frequency $\omega_{\perp}=2\pi\times 230~\mbox{Hz}\gg \omega_{||}$~\cite{cornell}. The transverse oscillator length is then $a_{\perp}=0.71~\mu\mbox{m}$ for $^{87}\mbox{Rb}$ atoms and $a_{\perp}=1.38~\mu\mbox{m}$ for $^{23}\mbox{Na}$ atoms. In what follows, we set $d=\pi$, which for our parameters corresponds to the lattice spacing of $2.23~\mu$m for $^{87}\rm Rb$ and $4.34~\mu$m for $^{23}\rm Na$ atoms.

Spinor condensates exhibit both spin-independent and spin-dependent nonlinear inter-atomic interactions, with the sign of the spin-dependent mean-field energy effectively determining the properties of the ground and excited states. A BEC of bosonic spin-1 atoms can be either {\em ferromagnetic} (e.g.~$^{87}{\rm Rb}$) or {\em polar} (e.g.~$^{23}{\rm Na}$) in nature~\cite{ho,ohmi}, with the two types of BECs having different average values of the total spin and different symmetry properties. The coupling coefficients for ``symmetric" spin-independent and ``antisymmetric" spin-dependent interaction terms in Eqs.(\ref{eq:TDGPEs}) are: 
\begin{align}
\lambda_s=\frac{2}{3}\left(\frac{a_0+2 a_2}{a_\perp}\right); \;\;\;\;\;\; \lambda_a=\frac{2}{3}\left(\frac{a_2-a_0}{a_\perp}\right).
\label{lambdas}
\end{align}
Here $a_0$ and $a_2$ characterise the s-wave scattering of two atoms in the channels with total spin equal to $0$ and $2$ respectively. The measured values of the scattering lengths for $^{87}\mbox{Rb}$ (see~table~\ref{tab:parameters}) result in $\lambda_a < 0$ \cite{kempen} and a {\em ferromagnetic} ground state of the condensate. In contrast, a $^{23}\mbox{Na}$ BEC has $\lambda_a > 0$ and the condensate is {\em polar}~\cite{klausen}. The strong inequality $|\lambda_a|\ll |\lambda_s|$, which holds for both types of the BECs, means that the spin-dependent interactions are much weaker than the spin-independent ones.

\begin{table}
\begin{center}
\begin{tabular}{||c|c|c|c|c||}
\cline{1-5}
	& $a_{0} [a_{B}]$ & $a_{2} [a_{B}]$ & $\lambda_{s}$ & $\lambda_{a}$ \\
\hline\hline
$^{87}\mbox{Rb}$ &  $101.8$ & $100.4$ & $1.49 \times 10^{-2}$ & $-6.94 \times 10^{-5}$\\
$^{23}\mbox{Na}$ & $50.0$ & $55.0$ & $4.08 \times 10^{-3}$ & $1.28 \times 10^{-4}$\\
\cline{1-5}
\end{tabular}
\end{center}
\caption{
Scattering lengths $a_{0}$, $a_{2}$ for $^{87}\mbox{Rb}$~\cite{kempen} and $^{23}\mbox{Na}$~\cite{klausen} atoms given in units of the Bohr radius $a_{B}$. The dimensionless 1D coupling constants $\lambda_{s}$, $\lambda_{a}$ used throughout the paper are obtained according to~\Eqs~(\ref{lambdas}) (see text).
}
\label{tab:parameters}
\end{table}

Under the assumption of a strong anisotropy and a tight confinement in the dimensions transverse to the optical lattice, the model~(\ref{eq:TDGPEs}) can be further reduced to a system of mean-field equations for the axial components of the wavefunctions. By introducing separable wavefunctions, $\Psi_j=\psi_j(x)\psi_\perp(y,z)$, with the transverse components $\psi_\perp(y,z)$ determined by the ground state of the tight harmonic trap~\cite{d_reduction}, and integrating~\Eqs~(\ref{eq:TDGPEs}) over the transverse directions $\{y,z\}$, the model for the spinor condensate is reduced to the 1D evolution equations for $\psi_{j}(x)$. The coupling constants are renormalised to values listed in~table~\ref{tab:parameters}. For later reference, we define the total density of atoms as $n_{\rm Tot} \equiv \Sigma \: n_j(x,t)=\Sigma \: |\psi_j(x,t)|^2$, where $n_{\pm 1,0}$ are the atomic densities of different hyperfine states.

The reduced dynamical system of 1D GP equations for $\psi_j(x)$ is characterised
by the conserved total number of atoms $\mathcal{N}=\int^{+\infty}_{-\infty} \! dx \left( n_{-1} + n_{+1} + n_{0} \right)$ and Hamiltonian (total energy of the system), $H = \int^{+\infty}_{-\infty}h(x) dx$,
with the Hamiltonian density:
\begin{align}
\label{energy}
h \; =& \sum_{j=0,\pm1} \! \left( \frac{1}{2} \; |\partial_{x} \psi_j|^2
+ V_{\rm OL} \; n_j + \frac{\lambda_s}{2} \; n_j^{2} \right)
\\
\nonumber
+ & \; \lambda_s \!\!\! \sum^{j\neq k}_{j,k=0,\pm1} \!\! n_{j}\: n_{k} + h_a.
\end{align}
The spin-dependent part of the Hamiltonian density, $h_a=(\lambda_a/2)\left| \vec{s} \right|^2$, 
is proportional to the square of the local spin expectation vector:
\begin{align}
\vec{s}\equiv \vec{\psi}^{\dag}\hat{\bf{F}}\vec{\psi}
= n_{\rm Tot} \vec{\zeta}^{\dag}\hat{\bf{F}}\vec{\zeta}.
\end{align}
Here $\vec{\psi}=(\psi_{+1},\psi_0,\psi_{-1})^{\rm T}$ is a three-component wave function, $\vec{\zeta}=(\zeta_{+1},\zeta_0,\zeta_{-1})^{\rm T}$ is the three component spinor, normalised according to the condition $\vec{\zeta}^{*T}\vec{\zeta}=1$, and $\hat{\bf{F}}=(\hat{F}_x,\hat{F}_y,\hat{F}_z)^{\rm T}$ is the vector of spin operators~\cite{ho,domain_th}, so that the local spin expectation vector yields:
\begin{align}
\vec{s} (x)\equiv
\left( \begin{array}{ccc}
\sqrt{2}\;\;{\mathcal Re}\{\psi^*_{0}\psi_{+1}+\psi^*_{-1}\psi_{0}\}\\
-\sqrt{2}\;\;{\mathcal Im}\{\psi^*_{0}\psi_{+1}+\psi^*_{-1}\psi_{0}\}\\
n_{-1}-n_{+1}
\end{array} \right).
\label{eq:spin3}
\end{align}
In what follows, we will only refer to the normalized spin expectation vector ({\em local spin})
\begin{align}
\vec{\mathcal S}\equiv \vec{s}/n_{\rm Tot},
\label{eq:spin}
\end{align}
where we note $\vec{\mathcal S} = (\mathcal{S}_x,\mathcal{S}_y,\mathcal{S}_z)^T$. Furthermore, we define the averaged spin expectation vector (i.e.~{\em total spin}) as: 
\begin{align}
\langle \vec{\mathcal S}\rangle\equiv \int \! dx \; \vec{s}(x) / \int \! dx \; |\psi(x)|^{2}.
\label{eq:avspin}
\end{align}
In the definition~(\ref{eq:avspin}) the average spin is normalized such that $\langle \vec{\mathcal S}\rangle$ takes the maximal value $1$. In the absence of a magnetic field, due to the rotational symmetry the total angular momentum of the system $\langle \vec{\mathcal S}\rangle$ is conserved. We note that $\langle \mathcal{S}_{z}\rangle=\int^{+\infty}_{-\infty} \left(n_{-1} - n_{+1} \right) dx$, known as the system's magnetisation, is the difference in the fraction of atoms in the Zeeman states $m_{F}=\pm1$. The conservation of the magnetisation in this system has a deep analogy in the optical parametric four-wave mixing process~\cite{fwm_yuri}. The four-wave mixing interaction, involving e.g.~the annihilation of a pair of atoms with $m_{F}=+1$ and $m_{F}=-1$, and creation of two atoms in the $m_{F}=0$ state, leads to phase conjugation in quantum optics and to matter-wave phase conjugation in BECs~\cite{goldstein:1999}.

\section{\label{sec:states}General properties of stationary states}
  
The stationary states of a spinor condensate can be found in the form $\psi_j(x,t)=\phi_j(x)\exp(-i\mu_jt)\exp(i\theta_j)$, where $\phi_j(x)$ are real wavefunctions, $\mu_j$ are chemical potentials, and $\theta_j$ are absolute phases of the three hyperfine components. As usual for physical systems with parametric interactions \cite{fwm_yuri}, the phase matching condition for the chemical potentials, $2\mu_0=\mu_{+1}+\mu_{-1}$, must be satisfied for a stationary state to exist. In addition, the relative phase between the components, $\Delta \theta= 2\theta_0-(\theta_{+1}+\theta_{-1})$, can only take values $0$ or $\pi$. The spatial profiles of the stationary state wavefunctions, $\phi_j(x)$, are then set by the following coupled time-independent 1D equations:  
\begin{align}
\label{eq:TIGPEs}
&         \mu_{-1} \phi_{-1}=
         \mathcal{L}_{\rm 1D}  \phi_{-1}
+ \lambda_a \left( \phi_{-1}^2 + \phi_{0}^2 - \phi_{+1}^2 \right) \phi_{-1}
\nonumber
\\
&
\;\;\;\;\;\;\;\;\;\;\;\;\;\;\;\; \pm \lambda_a \phi_{0}^2\phi_{+1}
\nonumber
\\
&         \mu_{0} \phi_{0}=
         \mathcal{L}_{\rm 1D} \phi_{0} 
+ \lambda_a \left( \phi_{-1}^2 + \phi_{+1}^2 \right) \phi_{0}
\\
&
\;\;\;\;\;\;\;\;\;\;\;\;\;\;\;\; \pm 2 \lambda_a \phi_{+1}\phi_{-1}\phi_{0} 
\nonumber
\\
&         \mu_{+1} \phi_{+1}=
         \mathcal{L}_{\rm 1D} \phi_{+1} 
+ \lambda_a \left( \phi_{+1}^2 + \phi_{0}^2 - \phi_{-1}^2 \right) \phi_{+1}
\nonumber
\\
&
\;\;\;\;\;\;\;\;\;\;\;\;\;\;\;\; \pm \lambda_a \phi_{0}^2\phi_{-1}
\nonumber
\end{align}
where $\mathcal{L}_{1D}=-\frac{1}{2}\frac{\partial^2}{\partial x^2}+ V_{\rm OL}(x) + \lambda_s \left( \phi_{-1}^2 + \phi_{0}^2 + \phi_{+1}^2 \right)$. The plus sign in front of the last term in~\Eqs~(\ref{eq:TIGPEs}) corresponds to stationary states with all three components {\em in-phase} with one another and therefore $\Delta \theta =0$. The minus sign is realized for $\Delta \theta =\pi$, and then one of the components ($m_{F}=+1$ or $m_{F}=-1$) of a stationary state is {\em out-of-phase} with the others. We note that {\em both} types of stationary states {\em always} exist in the system, regardless of the sign of $\lambda_a$. However, it is easy to see~\cite{domain_th}, that  the {\em in-phase} stationary states minimise the spin-dependent part of the total energy~(\ref{energy}) of the system, $H_{a} = \int^{+\infty}_{-\infty}h_{a}(x) dx$, for $\lambda_a<0$, while the {\em out-of-phase} stationary states minimise the spin-dependent energy for $\lambda_a>0$ \cite{ho}. The spin-dependent part of the Hamiltonian thus determines the {\em ground state} of the condensate: {\em ferromagnetic} for $\lambda_a<0$, and {\em polar} for $\lambda_a>0$~\cite{ho,domain_th,gu,sma_pu,sma_yi}.

For large atomic densities and atom numbers the spin-dependent interaction energy $H_{a}$ is sufficient to create spatial spin textures in lattice-free condensates and thus leads to peculiar miscibility properties of spinor BECs~\cite{strenger}. Due to the strong inequality of the coupling constants $|\lambda_a|\ll|\lambda_s|$, the spin-dependent interaction has little effect on the spatial density distributions of the stationary states for low and moderate peak densities. Hence in small atomic ensembles the internal spatial dynamics is suppressed~\cite{sma_pu,chapman_nature,sma_zhang}, although $H_{a}$ still determines the {\em phase structure} of the ground state and governs spin-mixing dynamics~\cite{sma_pu}. The spatial density structure of the condensate can be determined from Eqs.~(\ref{eq:TIGPEs}). 

When the size of a condensate is smaller than the spin healing length $\xi_{\rm spin}=2\pi/\sqrt{2|\lambda_{a}|n}$, the three components of the condensate can be treated as if trapped in the~{\em same} effective potential created by an optical lattice and the combined mean-field densities of the spinor components. This {\em single mode approximation} has been widely employed to describe the properties of the ground eigenstates of this effective potential~\cite{sma_pu,sma_zhang,sma_yi,robins}. Within this approximation, a stationary vector wavefunction of a spinor condensate can be represented in the form: $\phi_j(x) = \phi(x) \zeta_j$, and all the components share the common density profile~$\phi^{2}$ that obeys the single Gross-Pitaevskii equation:  
\begin{equation}\label{sma_eq}
         \mu \phi=
         -\frac{1}{2}\frac{\partial^2 \phi}{\partial x^2}+ V_{\rm OL}(x) \phi+ \lambda_s \phi^3,
\end{equation}
where $\mu\equiv\mu_0=\mu_{+1}=\mu_{-1}$. The spatial distribution of the local spin $\vec{\mathcal S}$ is governed by the relative distributions of condensate components. As a consequence of a shared density, the distribution of the local spin
is reduced to a single point in the 3D spin space.
 
Within the SMA, the {\em in-phase} (ferromagnetic) stationary states are characterised by the spinor~\cite{ho} :
\begin{align}
\label{hoferro}
\left( \begin{array}{ccc}
\zeta_{+1}\\
\zeta_{0}\\
\zeta_{-1}
\end{array} \right)=
\left(\! \begin{array}{ccc}
\cos^2 (\beta/2)\\
\sqrt{2} \cos (\beta/2) \sin (\beta/2)\\
\sin^2 (\beta/2)
\end{array}\!\right),
\end{align}
where $\beta$ is an arbitrary rotation angle around the $z$-axis (see \Fig~\ref{fig:sstates} for the definition). Accordingly, the squared expectation value of spin in this state is $\langle \vec{\mathcal S}\rangle^2 =1$. The spinor for {\em out-of-phase} (polar) stationary states is given by~\cite{ho}:
\begin{align}
\label{hopolar}
\left( \begin{array}{ccc}
\zeta_{+1}\\
\zeta_{0}\\
\zeta_{-1}
\end{array} \right)=
\left( \begin{array}{ccc}
-\frac{1}{\sqrt{2}} \sin \beta\\
\cos \beta\\
\frac{1}{\sqrt{2}} \sin \beta
\end{array} \right),
\end{align}
and $\langle \vec{\mathcal S}\rangle^2 =0$ for this state. Our numerical solutions of Eqs.~(\ref{eq:TIGPEs}), which are presented in sections~\ref{sec:bloch}, \ref{sec:solitons} and~\ref{sec:sts}, confirm the validity of the SMA for a variety of extended and spatially localized nonlinear stationary states (of both polar and ferromagnetic type) for spinor BECs in optical lattices. However, as will be shown below, there exist nonlinear stationary states that cannot be described by the SMA.

\section{\label{sec:bloch}Stability of spinor Bloch waves}

When the interatomic interaction can be neglected the eigenstates of~\Eqs~(\ref{eq:TIGPEs}) take the form of linear Bloch states: 
$\phi(x)\sim \phi_{k}(x) {\rm exp}(ikx)$, where $\phi_{k}(x)=\phi_{k}(x+d)$ has the periodicity of the lattice, and the wave number $k$ is selected from the first Brillouin zone (BZ), i.e.~$-1<k<1$ in our units. Then the linear spectrum $\mu(k)$ of the Bloch states has a characteristic band-gap structure with the lowest two bands shown in~\Fig~\ref{fig:bgd}. In the presence of interactions, the most fundamental stationary states of a BEC in an optical lattice are the periodic, spatially extended nonlinear Bloch waves~\cite{pearl} that transform into linear Bloch waves for sufficiently low atomic densities (nonlinearity).

It is known that Bloch states of a single-component BEC with repulsive atomic interactions, corresponding to the top edge of the ground band (i.e. the edge of the first Brillouin zone), exhibit complex nonlinear dynamics due to modulational instability (MI) ~\cite{latticemi}. This instability arises because the anomalous diffraction at the band edge effectively changes the sign of the atomic interactions and can lead to effects that are normally absent in a repulsive condensate, such as formation of spatially localized (bright) gap solitons. As a result of such instability development, gap solitons trains can form in a process of condensate evolution at the edge of a Brilllouin zone ~\cite{latticemi,beatka1,beatka2}.

\begin{figure}[ht]
\includegraphics[width=10\textwidth,height=7cm,keepaspectratio]{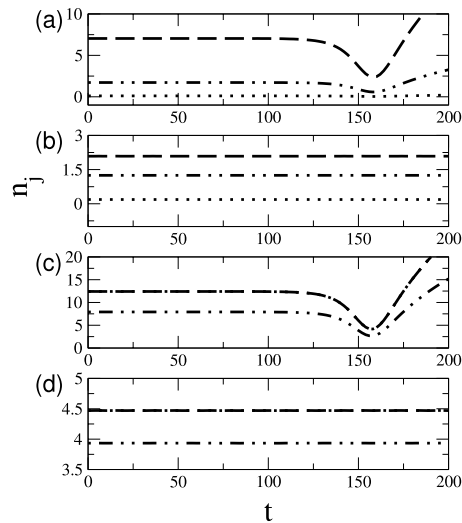}
\caption{\label{fig:mi} 
Development of modulation instability for $F=1$ spinor Bloch states in (a) ferromagnetic ($^{87}\rm Rb$) and (c) polar ($^{23}\rm Na$) BECs at the top edge of the first band ($V_{0}=4$, $\mu=1.4$). Shown are the peak densities of the $m_{F}=+1$ (dashed line), $m_{F}=0$ (dash-dotted line) and $m_{F}=-1$ (dotted line) condensate components at $x=100$. For comparison, the behaviour of the modulationally stable ground state nonlinear Bloch waves at $\mu=1.28$ are shown for a (b) ferromagnetic ($^{87}\rm Rb$) and (d) polar ($^{23}\rm Na$) BECs.
}
\end{figure}

In the absense of a lattice, the stability properties of a spinor BEC confined in a harmonic trap are well established. Namely, a condensate with ferromagnetic type of spin-dependent atomic interactions  is found to be modulationally unstable, and the polar BEC is stable ~\cite{mi,robins}. Here, by performing nymerical analysis of condensate evolution~\footnote{The dynamics of the Bloch states is investigated by solving~\Eqs~(\ref{eq:TDGPEs}) using an adaptive Runge-Kutta-Fehlberg method within the high level programming language XMDS~(http://www.xmds.org). We perform reliable simulations of one-dimensional condensate dynamics with $128$ grid points per lattice well on an $x$-range of $2\pi \times 128$ lattice sites. This corresponds to a grid spacing $\Delta x$ of $0.0491$, approximately 40 times smaller than the condensate healing length $\xi=\left( 8 \pi n_{\rm Tot} a_{s} \right)^{-1/2}$ for the BEC Bloch waves.}, we find that Bloch states corresponding to the upper edge of the ground band [see Fig. 1(c)] for both, ferromagnetic ($^{87}\rm Rb$) and polar ($^{23}\rm Na$) BECs are modulationally unstable. In contrast, ground Bloch states corresponding the lower edge of the band [see Fig. 1(d)] for both $^{87}\rm Rb$ and $^{23}\rm Na$ are modulationally stable, irrespective of the phase relationahsip between the hyperfine components.

Typical results of our numerical simulations demonstrating the stable and unstable dynamics of Bloch states in an optical lattice of moderate depth are presented in~\Fig~\ref{fig:mi}. For the chosen parameters, the duration of evolution sufficient to detect the onset of MI is $200$ time units ($140$ms). The onset of the MI occurs at $t\sim130$ ($90$~ms) for the unstable states of both ferromagnetic and polar condensate types at the top of the first band, as seen in \Fig~\ref{fig:mi}(a,c). The time scale of the MI development is comparable with the time scale associated with density-dependent interactions, $t_{s}=\lambda_{s}n_{\rm Tot}$. The peak densities of the initial states $n_{\rm Tot}=8.9$ for $^{87}\rm Rb$ and $n_{\rm Tot}=32.7$ for $^{23}\rm Na$ give $t_{s}=0.134$ ($92.5~\mu$s) for rubidium and $t_{s}=0.133$ ($92~\mu$s) for sodium.  For the same lattice parameters, Bloch states from the bottom of the BZ demonstrate stable dynamics, as shown in \Fig~\ref{fig:mi}(b,d).

\begin{figure}
\includegraphics[width=10\textwidth,height=6.5cm,keepaspectratio]{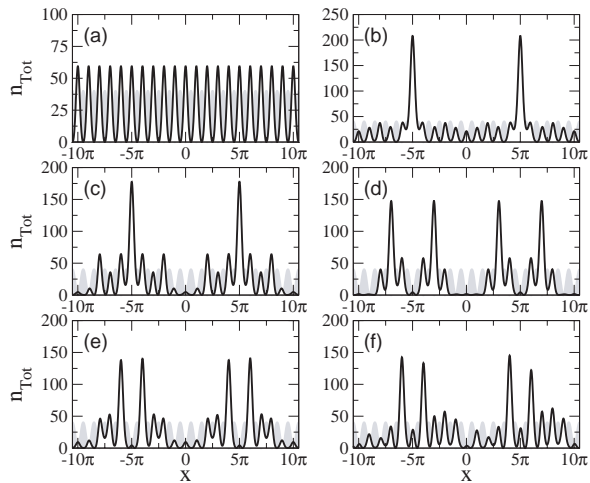}
\caption{\label{fig:mispin} 
Instability-induced dynamics for a $^{23}\rm Na$ condensate at the top edge of the ground band ($\mu=1.12, V_{0}=2$). Shown are snapshots of the total BEC density (solid lines) with the periodic OL potential (gray shading, magnified by a facto of 20) at the times (a) $t=0$, (b) $t=20$, (c) $t=40$, (d) $t=100$, (e) $t=120$, and (f) $t=200$.}
\end{figure}

The dynamical instability of the Bloch states at the band edge underpins the existence of the spatially localized states. In order to further investigate the dynamics of the Bloch states beyond the onset of the MI and initiate formation of localized structures~\cite{beatka2}, we introduce initial periodic phase modulation, $\exp\{i\cos(\delta x)\}$, imprinted onto all the hyperfine components in the initial state. As an initial state, we take an unstable Bloch wave corresponding to the top of the first band in a relatively shallow ($V_{0}=2$) lattice. This results in larger peak densities $n_{\rm Tot}$ of $16.6$ for $^{87}\rm Rb$ and $61.7$ for $^{23}\rm Na$, and shorter interaction time scales $t_{s}\sim0.25$ ($0.2$~ms) for both atomic species. In~\Fig~\ref{fig:mispin} we show snapshots of the total density $n_{\rm Tot}$ in a shallow lattice for anti-ferromagnetic matter waves. The wavelength of the phase modulation, $\lambda=2\pi/\delta$, corresponds to ten lattice sites for $\delta = 0.2$. As a result of the phase imprinting, strong density modulations develop at the initial stages of the evolution. Periodic density modulations trigger modulational instability and greatly accelerate the process of the condensate localization, as demonstrated previously for a single-component BEC \cite{beatka2}.

The scenario of dynamical localization is almost identical for both $^{87}\rm Rb$ and  $^{23}\rm Na$ condensates (shown in ~\Fig~\ref{fig:mispin} for a sodium BEC). This dynamics is primarily determined by the spin-independent interactions and hence is similar to the dynamics of a single component condensate with repulsive inter-atomic interactions \cite{beatka1}. The development of the localized structures is complex, with a transition from single-peak soliton-like structures [Fig. \ref{fig:mispin} (b)] to unstable ``bound states" of two neighbouring solitons [Fig. \ref{fig:mispin} (d)]. The nature of the spin-dependent interactions turns out to play a role in the interaction of localized structures. Namely, for polar ($^{23}\rm Na$) condensate the neighbouring localized states tend to repel each other due to the presence of {\em out-of-phase} components. Conversely, for a ferromagnetic ($^{87}\rm Rb$) condensate, the neighbouring solitons with all three components {\em in-phase} attract each other. In general the ``bound states" do not persist over long evolution times and the dynamically formed solitonic structures eventually disappear [\Fig~\ref{fig:mispin}(f)].

\section{\label{sec:solitons}Spinor gap solitons}
\begin{figure}
\includegraphics[width=\columnwidth,keepaspectratio]{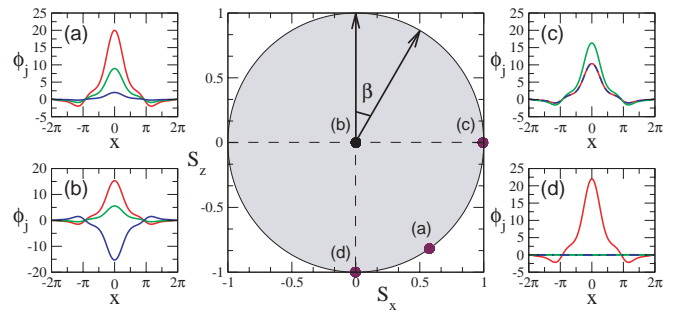}
\caption{\label{fig:sstates} 
Examples of spatial structure of the (a,c,d) ferromagnetic-type (\emph{in-phase}) and (b) polar-type, (\emph{out-of-phase}) gap solitons of a {\em polar} $^{23}\rm Na$ condensate in an optical lattice ($V_{0}=4, \mu=2.8$). Red, green, and blue lines show the $\phi_{-1}$, $\phi_{0}$, and $\phi_{+1}$ components, respectively. The corresponding values of the local spin vector $\vec{\mathcal S}=({\mathcal S}_x,{\mathcal S}_y,{\mathcal S}_z)$ are plotted in the plane ${\mathcal S}_y=0$ of the 3D spin sphere (shaded); $\beta$ is the Euler rotation angle.
}
\end{figure}
Unlike the spatially extended nonlinear Bloch waves, spatially localized states of a condensate in an optical lattice can exist only within the spectral band gap. Here we describe the properties of stationary \emph{three-component (vector) gap solitons}. To this end, we numerically solve time-independent equations~(\ref{eq:TIGPEs}), for both types of spin-dependent nonlinear interactions, looking for spatially localized solutions within the gap. The results of our findings are summarised in~\Fig~\ref{fig:sstates}, where we plot the local spin vector in the $\{\mathcal{S}_x,\mathcal{S}_z\}$ coordinate plane ($\mathcal{S}_y=0$ due to the phase matching condition for stationary states). As shown in the figure, all stationary three-component gap solitons can be divided into the two types, {\em in-phase} (ferromagnetic-type) and {\em out-of-phase} (polar-type) states. We stress that {\em both} types of solution exist for {\em both} ferromagnetic ($\lambda_a<0$) and polar ($\lambda_a>0$) types of spin-dependent interactions. In agreement with the single mode approximation, all the in-phase states can be described as a point on a circle of unit radius on the $\{\mathcal{S}_x,\mathcal{S}_z\}$ coordinate plane. The peak density of the $m_{F}=0$ component ranges from~$0$ at point~(d), to half of the total peak density, $n_{\rm Tot}/2$, at point~(c). In contrast, all the out-of-phase gap states correspond to the point at the origin $\{\mathcal{S}_x,\mathcal{S}_z\}=\{0,0\}$. 

Profiles corresponding to point~(a) in~\Fig~\ref{fig:sstates} represent a typical three-component in-phase gap soliton, where the strong localization occurring in a nonzero spin component, e.g.~$m_{F}=-1$, induces the localization in the other two hyperfine states. In contrast, for the in-phase gap solitons corresponding to $\mathcal{S}_z=0$ and represented by point~(c), the amplitudes of the component wavefunctions for $m_{F}=+1$ and $m_{F}=-1$ are equal, see~\Fig~\ref{fig:sstates}(c), so that the populations of the states are: $n_{0}=n_{\rm Tot}/2$ and $n_{-1}=n_{+1}=n_{\rm Tot}/4$. Two of the spinor components are equal: $\zeta_{+1}=\zeta_{-1}$. At point~(d) (or its diametrically opposite), $\zeta_0=0$ and two spin components have zero amplitudes to yield a maximum (or minimum) value of the magnetisation. The example in~\Fig~\ref{fig:sstates}(d) shows a gap soliton with all atoms in the $m_{F}=-1$ state, which maximises the magnetisation.

All the out-of-phase solutions to~\Eqs~(\ref{eq:TIGPEs}) have $\phi_{+1}$ and $\phi_{-1}$ components that are equal in their absolute values but $\pi$ out of phase, which effectively leads to a cancellation of the spin interaction terms in model~(\ref{eq:TIGPEs}) and a zero value of the magnetisation. It should be noted that some states represented by points~(b) and the in-phase state indicated by point~(c) in~\Fig~\ref{fig:sstates} can not be distinguished in an experiment by simple reconstruction of the component density profiles, as they might differ from each other only by a relative phase. The recently developed nondestructive method of polarization-dependent phase-contrast imaging~\cite{higbie} opens up new possibilities for probing spinor BECs, and can ultimately enable the practical identification of the in-phase and out-of-phase states and lead to recognition of ground states and long-lived metastable spin states.

\begin{figure}
\includegraphics[width=10\textwidth,height=3.5cm,keepaspectratio]{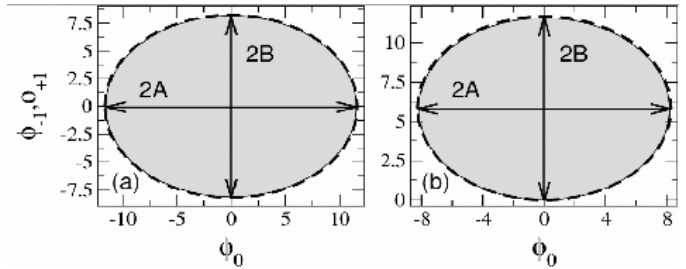}
\caption{\label{fig:maxima} 
Comparison of the values of gap soliton peak amplitudes, $\phi_{+1}$ and $\phi_{-1}$ at $x=0$ (dashed lines), with the amplitudes of a homogeneous condensate (solid lines) for (a) out-of-phase and (b) in-phase stationary states of a $^{87}$Rb condensate in a lattice with $V_0=4$, plotted as a function of $\phi_{0}$. Gap soliton is taken at $\mu=2.8$, and the homogeneous condensate at $\mu_{\rm eff}=2.0$, to account for the shift of the spectrum from zero (see text). Analytically obtained parameters of the semi-axes of the ellipses and the ellipse displacement are (a) $A=11.59$, $B=8.19$ and (b) $A=8.21$, $B=5.8$ and $\Delta_{+1}=B$.
}
\end{figure}

In general, stationary gap solitons of the in-phase type can have atoms populating one or three hyperfine states, while the solitons of the out-of-phase type can also form two component states. In \Fig~\ref{fig:maxima} we show the variation of the soliton peak amplitudes in different hyperfine states, depending on a given peak amplitude in the $m_{F}=0$ state, for a particular set of lattice parameters and a chemical potential within the first gap. We compare these with the peak amplitudes calculated for a homogeneous condensate. The introduction of a periodic potential shifts the chemical potential of the extended solution with the lowest chemical potential (the bottom of the first band) relative to the homogeneous condensate (the solution at $V_0 = 0$), as can be seen in~\Fig~\ref{fig:bgd}(a,b).  Thus in comparing peak amplitudes it is important that the same effective chemical potential is used.  To this end we use for the homogeneous condensate $\mu_{\rm eff} = \mu - \mu_0$ where $\mu_0$ is the chemical potential of the bottom of the first band.

Considering the {\em out-of-phase} states in a homogeneous case we set, $f=\sqrt{2}\phi_{+1}=-\sqrt{2}\phi_{-1}$ and $g=\phi_{0}$, so that~\Eqs~(\ref{eq:TIGPEs}) simplify to:
\begin{align}
\label{eq:PolarGPEs}
\mu_{\rm eff} \; f =  \lambda_a \left( f^{2} + g^{2} \right) f;
\;\;\;\;\;\;\;
\mu_{\rm eff} \; g = \lambda_a \left( f^{2} + g^{2} \right) g.
\end{align}
This set of equations has an implicit analytical solution which can be written in the form of an ellipse equation for $\phi_{j}$:
\begin{align}
\frac{\phi_{0}^{2}}{A^{2}} +\frac{\phi_{\pm 1}^{2}}{B^{2}} =1,
\end{align}
where $A=\sqrt{\mu_{\rm eff}/\lambda_{s}}$ and $B=\sqrt{\mu_{\rm eff}/(2\lambda_{s})}$ are the semi-major and the semi-minor axes of the ellipse respectively. For the amplitudes of an {\em in-phase} state, equations~(\ref{eq:TIGPEs}) cannot be simplified, and give the implicit solution:
\begin{align}
\frac{\phi_{0}^{2}}{A^{2}} +\frac{(\phi_{\pm 1}-\Delta_{+1})^{2}}{B^{2}} =1.
\end{align}
Here the ellipse parameters $A$, $B$, and the displacement of the ellipse centre $\Delta_{+1}$ are defined as:
$A^{2}=\left[ 2(\lambda_{s}+\lambda_{a})/\mu_{\rm eff} \right]^{-1}$,
$B^{2}=\left[ 4(\lambda_{s}+\lambda_{a})/\mu_{\rm eff} \right]^{-1}$ and
$\Delta_{+1}=\frac{1}{2}\sqrt{\mu_{\rm eff}/(\lambda_{s}+\lambda_{a})}$.
These solutions, derived from the lattice-free equations, predict the amplitudes of the multi-component gap solitons with good accuracy (see \Fig~\ref{fig:maxima}).

As both the in-phase and out-of-phase types of states exist in any spin-1 BEC, the generation of both classes of gap solitons could be experimentally feasible. A small atom number BEC cloud of optically trapped atoms can initially be transferred to the desired $m_{F}=+1$ or $m_{F}=-1$ spin state by Landau-Zener rf-sweeps~\cite{ketterle}. Then a single-component gap soliton (corresponding to point~(d) in~\Fig~\ref{fig:sstates}) in an optical lattice can be formed~\cite{gap_soliton}. Finally, by applying a magnetic field perpendicular to the $z$-axis, a spin-rotation can be induced to change relative population of the spin components within a stationary state.

\begin{figure}[t]
\includegraphics[width=10\textwidth,height=3cm,keepaspectratio]{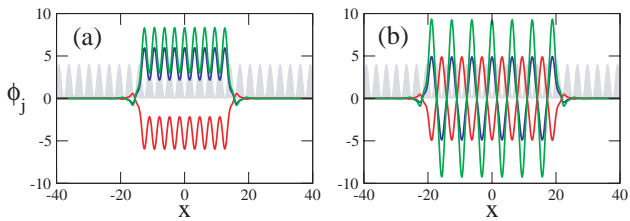}
\caption{\label{fig:gapwave} 
Self-trapped gap states at  $\mu = 2.8$ corresponding to the nonlinear Bloch waves of a $^{87}$Rb condensate (a) in the ground state and (b) at the edge of the first band of the optical lattice potential (shaded) with $V_0 = 4$. Red, green, and blue lines show $\phi_{-1}$, $\phi_{0}$ and $\phi_{+1}$ hyperfine states, respectively.
}
\end{figure}
\section{\label{sec:sts}Self-trapped states}
A second type of spatially localized states existing in a lattice, the ``self-trapped'' states, have recently been investigated in a single-component condensate~\cite{st_exp,selftrapped,st_theory}. These states are nonlinear Bloch waves which at sufficient densities experience de-tuning of the chemical potential deep into the linear band gap. Consequently they may be localized by Bragg reflection and ``truncated'' spatially. The localization has the same physical origin as for the gap solitons, and therefore these states connect the two fundamental types of nonlinear lattice states, the extended nonlinear Bloch waves and the localized gap solitons. The truncated nonlinear Bloch waves may be originate from any periodic Bloch state, such as e.g.~the ground state nonlinear Bloch wave. 

Here we generalize this concept to the multi-component Bose-Einstein condensates and demonstrate the existence of \emph{spinor gap waves} (see~\Fig~\ref{fig:gapwave}). Akin to a single component self-trapped state, the ground state spinor gap waves only exist beyond a critical lattice depth, sufficient for localization to take place. Beyond this point the gap waves may be localized with an arbitrary spatial extent~\cite{selftrapped}. Nonetheless, in contrast to a single component BEC the spin degree of freedom results in possibilities for different relative populations of the hyperfine components, similar to the case of spinor gap solitons (as shown in~\Fig~\ref{fig:sstates}). In addition, they also form in-phase and out-of-phase spin configurations for any type of spin-dependent interactions. Examples of out-of-phase truncated nonlinear Bloch waves for $^{87}$Rb are shown in~\Fig~\ref{fig:gapwave}.

We note, that the spatial structure of self-trapped gap waves can be adequately described within the single-mode approximation, and hence Eqs. (\ref{hoferro}) and (\ref{hopolar}) determine relative fractional populations for different orientations of spin in a self-trapped state.

\section{\label{sec:nonsma}Beyond the Single Mode Approximation}
\begin{figure}[t]
\includegraphics[width=10\textwidth,height=3cm,keepaspectratio]{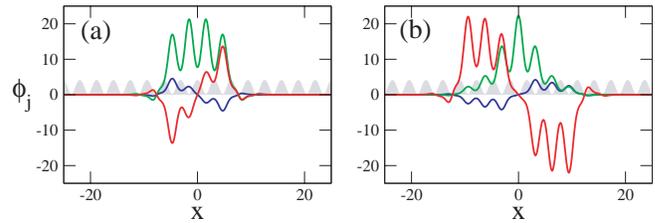}
\caption{\label{fig:nsma1} 
Non-SMA stationary states for a $^{23}{\rm Na}$ condensate: (a) centred on the potential maximum (i.e.~\emph{off-site}) and (b) centred on the potential minimum (i.e.~\emph{on-site}). Both stationary states are found for the values of the chemical potentials: $\mu_{+1}=2.8$, $\mu_{-1}=2.7$ and $\mu_{0}=\frac{1}{2}(\mu_{+1}+\mu_{-1})=2.75$. Red, blue, green lines mark $\phi_{+1}$, $\phi_{-1}$ and $\phi_{0}$ components, respectively. Shaed areas show the corresponding optical lattice potential ($V_0=4$).
}
\end{figure}
Apart from the ``fundamental'' gap states considered above, and adequately described by the single-mode approximation, an optical lattice supports the existence of other spinor gap states with more complex spatial structure.
Under special conditions, {\em off-site} and {\em on-site} gap solitons that are extended over multiple lattice sites and centered on a maximum or minimum of the lattice potential respectively can be found. These and other ``excited'' gap states contain a larger number of atoms than the fundamental solitons. Most importantly, due to the dramatic differences between the spatial density distributions in the three hyperfine states, they do not obey the single mode approximation. Examples of non-SMA states obtained from equations~(\ref{eq:TIGPEs}) for a $^{23}$Na condensate are shown in~\Fig~\ref{fig:nsma1}. 

The existence of the states that do not obey the SMA is counter-inituitive as the size of gap solitons is an order of magnitude smaller than a typical spin domain~\cite{ketterle}. These states may be regarded as a composite of the self-trapped states {\em and} gap solitons, in which a gap soliton is localized in the combined effective potential of the self-trapped state and the lattice. Straightforwardly, the localized structure in $m_{F}=0$ is induced by a $\pi$ phase jump in the density envelope of a strongly populated component, either $m_{F}=-1$ or $m_{F}=+1$. Importantly, here the spin degeneracy has been removed due to the difference in chemical potentials, i.e.~$\mu_{-1}\neq\mu_{+1}$, so the populations in each hyperfine component are no longer governed by simple rotation through spin space [see Eqs. (\ref{hoferro}) and (\ref{hopolar})]. Furthermore, it is clear that within these states one component is weakly populated, leading to an effective two-component system. Despite this reduction in the number of the hyperfine states, the spin becomes here an important quantity, with the solutions possessing a complex spin structure.
\begin{figure}
\includegraphics[width=\columnwidth,keepaspectratio]{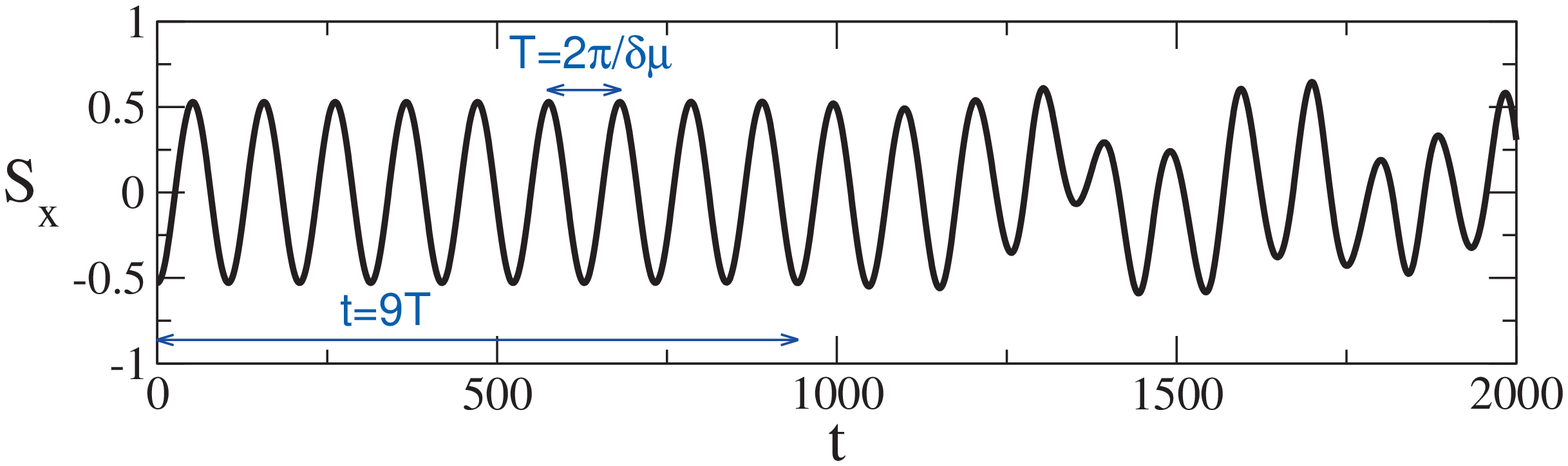}
\caption{\label{fig:decayspin} 
Spin dynamics of unstable off-site non-SMA state of $^{23}{\rm Na}$ for parameters given in~Fig~\ref{fig:decay1}. The magnitude of $\mathcal{S}_{x}$ is taken at $x=\frac{3}{2}\pi$. Initially, the local spin oscillates with period $T=\frac{2 \pi}{\delta \mu}\simeq 104.7$. For a time interval of nine oscillations, which correspond to $t=942.5$, the dynamics is stable. Afterwards the unstable evolution commences. 
}
\end{figure}

The spatially separated wavefunctions of the different hyperfine components give $\nabla \vec{\zeta} \neq 0$ and lead to a violation of the single mode approximation. Hence the spin-dependent part of the total energy $H_{a}$ is no longer constant. As an effect the values of the local spin tend to cover the whole range $0<|\vec{\mathcal{S}}|<1$ and gap solitons acquire spin structure across the localization region, as shown in~\Fig~\ref{fig:spinproj}. 

By numerical simulations we determine that the on-site non-SMA soliton shown in~\Fig~\ref{fig:nsma1}(b) is stable. However, due to the difference in the chemical potential between the components the local spin exhibits spin precession dynamics as given by the expression:
\begin{align}
\vec{\mathcal S} (x,t)=
\left( \begin{array}{ccc}
\mathcal{S}_x(x,0)
\cos{(\delta \! \mu \: t + \delta \theta)}\\
\mathcal{S}_y(x,0)
\sin{(\delta \! \mu \: t + \delta \theta)}
\\
\mathcal{S}_z(x,0)
\end{array} \right),
\end{align}
where $\delta\theta \equiv \frac{1}{2}(\theta_{-1}-\theta_{+1})$ and $\delta\!\mu \equiv \frac{1}{2}(\mu_{-1}-\mu_{+1})$ according to the phase-matching conditions for stationary states. For real stationary states the local spin oscillates with the frequency $\omega=\delta\!\mu$. As the non-SMA states are made of ``twisted mode'' gap waves in $m_{F}=\pm1$ components which are $\pi$-out of phase from each other as shown in~\Fig~\ref{fig:nsma1}, $\delta\theta=\pi$. We have checked that in the regimes of their stability the non-SMA states do indeed exhibit spin oscillations at the frequency $\delta\!\mu$. 
We display those in~\Fig~\ref{fig:decayspin} showing $\mathcal{S}_{x}$ for the unstable state in~\Fig~\ref{fig:nsma1}(a).
Due to the precession of the local spin in the $\{\mathcal{S}_{x},\mathcal{S}_{y}\}$ plane, each point of the soliton behaves as if it was exhibiting a permanent magnetic field along the $z$-axis. Therefore the non-SMA solitons can be thought of as being \emph{self-magnetised}. It is conceivable that polarization-dependent phase-contrast imaging~\cite{higbie} could resolve this effect. We note that although the {\em local} spin exhibits precession around the $\mathcal{S}_{z}$ axis, the total angular momentum is conserved. 
\begin{figure}
\includegraphics[width=10\textwidth,height=6.5cm,keepaspectratio]{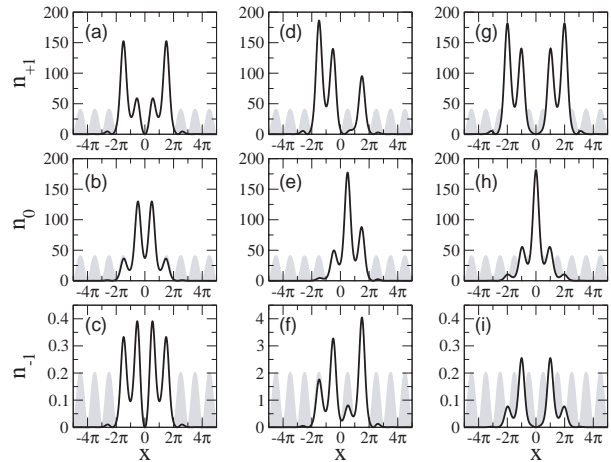}
\caption{\label{fig:decay1} 
Non-SMA stationary states of $^{23}{\rm Na}$ for parameters: $\mu_{+1}=1.28$, $\mu_{-1}=1.4$ and $\mu_{0}=\frac{1}{2}(\mu_{+1}+\mu_{-1})=1.34$ and $V_{0}=2$. All hyperfine components of the unstable off-site soliton at $t=0$ (a,b,c) and at $t=2000$ (d,e,f) are compared against the corresponding profiles of a stable on-site non-SMA state (g,h,i).
}
\end{figure}

Although the on-site state is stable, the off-site state shown in~\Fig~\ref{fig:nsma1}(a) is unstable and its stability properties depend on the depth of the lattice potential and the degree of localization due to the chemical potential. For the chosen parameters, this state decays after nine periods of local spin oscillation. Due to the self-trapping effect, this unstable off-site state does not lose atoms by radiation but instead starts to exhibit spin interaction induced Josephson-like oscillations \cite{Salerno_J} between the left-hand side and the right-hand side potential wells. 
To gain some insight into the decay mechanism of the unstable state, we analyse the development of the instability. Accordingly, we plot in~\Fig~\ref{fig:decay1} profiles of different $m_{F}$ components of the off-site state at $t=0$ (left column) and $t=2000$ (middle column) corresponding to the initial and final time samples of our simulations. The matching spin projections at these times are presented in~\Fig~\ref{fig:spinproj}. We compare the unstable off-site state against the stable on-site non-SMA state (right column in both~\Fig~\ref{fig:decay1} and~\Fig~\ref{fig:spinproj}). In the course of the unstable evolution, the off-site state transforms into an on-site state and then evolves into an unstable ``bound-state'', similar to those emerging from the MI-induced dynamics (see section~\ref{sec:bloch}). 

\begin{figure}
\includegraphics[width=10\textwidth,height=4.5cm,keepaspectratio]{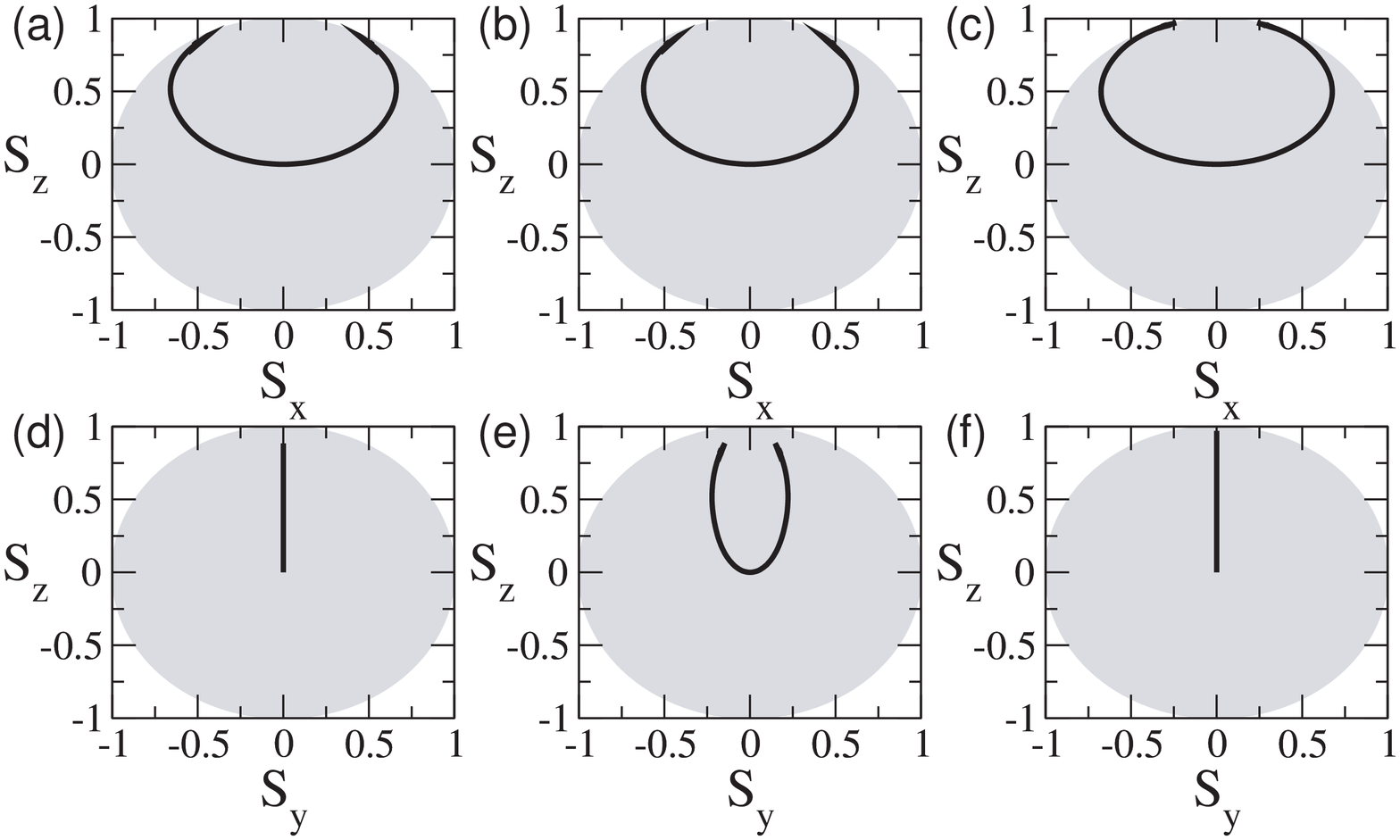}
\caption{\label{fig:spinproj} 
Projections of local spin of the non-SMA states of $^{23}{\rm Na}$ (shown  in~\Fig~\ref{fig:decay1}) onto the planes $S_{y}=0$ (a,b,c) and $S_{x}=0$ (d,e,f) of the spin sphere (shaded). Distribution of the local spin vector of the unstable off-site non-SMA state in its (a,d) initial (a,d) and (b,e) final, i.e.~at~$t=2000$ configurations. For comparison, panels (c,f) show the spin distribution for the stable on-site state. 
}
\end{figure}

\section{\label{sec:dynamics}Non-stationary dynamics}
If a spinor BEC is initially prepared in a non-stationary spin configuration, a general spin-mixing dynamics occurs~\cite{sma_pu,chapman,sma_zhang,chapman_nature}, i.e.~exchange of population between the central and sideband fields. The role of the ferromagnetic (or polar) interactions in this dynamics is to raise (lower) the energy of the $m_{F}=0$ component relative to the energies of the other two components, until the system reaches its ferromagnetic or (polar) ground state. Predicted theoretically~\cite{pu_quantum_mixing,pu_mixing}, the spin-mixing dynamics of large-size harmonically trapped condensates has recently been experimentally demonstrated~\cite{chapman,sengstock}. Unfortunately, the spin-mixing oscillations were observed for only a few periods, due to the damping by fragmentation of the spinor BEC~\cite{chapman,sengstock}. 

In the framework of the mean-field theory, the coherent evolution of spinor components does not lead to damping of spin oscillations. As the size of a single gap soliton is much smaller than the spin healing length, in contrast to larger condensate clouds~\cite{chapman,sengstock}, their evolution can be very accurately described by the single mode approximation. The conservative non-equilibrium dynamics within the SMA generally consists of periodic exchange of populations between different hyperfine components, which results in the dynamics of spinors: $\zeta_j(t)=\sqrt{n_j(t)}\exp(-i \theta_j(t))$~\cite{chapman_nature}. Within the SMA the spin-dependent energy landscape is governed by the Hamiltonian~\cite{sma_zhang}:
\begin{align}
\label{eq:landscape}
\mathcal{H} = \gamma \rho_{0} \left( 1 - \rho_{0} 
+ \sqrt{(1-\rho_{0})^{2} - m^{2}} \right) \cos(\Delta \theta)+ \frac{\gamma}{2} m^{2}
\end{align}
with $\rho_0=n_{0}/n_{\rm Tot}$ being the relative population of the $m_{F}=0$ state. Here $m=n_{+1}-n_{-1}$ is the value of (the local) magnetisation and $\gamma = \lambda_a \; \mathcal{N} \! \int \! dx \: (n_{\rm Tot})^{2}$ is the coupling. $\Delta \theta$ is the relative phase of the condensate, as defined in section~\ref{sec:states}. These values are extracted from the multi-component gap soliton. 

We note here that there are two ways to initiate the spin-mixing dynamics. The first one is via a non-stationary relative phase~\cite{pu_quantum_mixing,Lewenstein_spinor}. The second is by means of population imbalance. Both these procedures give equivalent results. It has previously been shown~\cite{pu_quantum_mixing,pu_mixing} that if the population is averaged over many realizations with different values of the initial condensate phase, the spin-mixing dynamics can be halted in the averaged picture, and the system may reach a steady-state.

\begin{figure}
\includegraphics[width=10\textwidth,height=3.5cm,keepaspectratio]{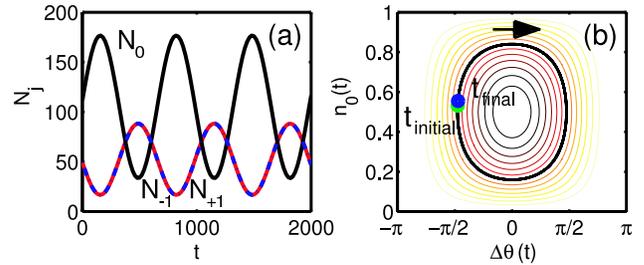}
\caption{\label{fig:noneq}
Dynamics of non-equilibrium states in~$^{87}{\rm Rb}$ ($\mu=2.8$, $V_{0}=4$). (a) Spin mixing oscillations of the populations between $m_{F}=-1$ (red), $m_{F}=+1$ (blue, dashed-dotted) and $m_{F}=0$ (black) hyperfine components. (b) The dynamical trajectories in the $n_0(\Delta \theta)$ phase space for the scenario presented in panel~(a). Initial (final) location in the phase space is marked by the green (blue) point. The arrow indicates the direction of the spin mixing oscillations in the phase space.
}
\end{figure}

We monitor the relative phases of solitons with the introduced population imbalance and plot their trajectories in the $n_{0}(\Delta \theta)$ phase space, as shown in~\Fig~\ref{fig:noneq}. In the figure we also compare the resulting trajectories with the energy landscapes, \Eq~(\ref{eq:landscape}), determined separately for each scenario. We prepare initial states by shifting the relative phases or by introducing a population imbalance in a stationary gap soliton, e.g.~as shown in~\Fig~\ref{fig:sstates}. An example of the dynamics in the first case is shown in~\Fig~\ref{fig:noneq}, with the green point in~\Fig~\ref{fig:noneq}(b) representing the position of the initial state. In the latter case we generate the SMA states under $0<|\vec{\mathcal{S}}|<1$ by parametric transfer of the population from $m_{F}=\pm1$ to $m_{F}=0$ from the in-phase stationary state represented by point~(c) in~\Fig~\ref{fig:sstates}. Our procedure yields: $n_{0} > n_{\rm Tot}/2$ and $n_{-1}=n_{+1} < n_{\rm Tot}/4$, which is equivalent to moving upwards in the diagram~\Fig~\ref{fig:noneq}. This situates the local spin of the engineered state inside of the spin sphere, at the ``equator'', between point~(c) and point~(b) in~\Fig~\ref{fig:sstates}. Both of the above procedures result in initial states that exhibit spin-mixing dynamics along closed phase space trajectories. We emphasize that not only the amplitude but, most importantly, also the period of the spin-mixing oscillations depends on the degree of the population imbalance, i.e.~the larger the imbalance, the longer the period of the oscillations (cf. \cite{chapman}).

In order to model the process of controlled manipulation of the coherent spinor dynamics, we simulate condensate dynamics under the influence of a pulsed magnetic field, as described in Ref.~\cite{chapman}. As we show below, the resulting management of the relative phase due to the quadratic Zeeman effect can lead to the arrest of the spin-mixing oscillations in the spatially localized state of the spinor condensate described by the SMA. As this effect is reversible, it allows flexible manipulations of the multi-component gap solitons.

The application of the magneticthe external magnetic field ${\bf B}$ along the $z$-axis results in a Zeeman shift of hyperfine states:
\begin{align}
\label{eq:ZeemanGPEs}
 i \frac {\partial \Psi_{\pm 1}}{\partial t} =&
         \;	\mathcal{L}_{1D}  \Psi_{\pm 1} + \varepsilon_{\pm 1} \Psi_{\pm 1}
          \nonumber
          \\
          &
         + \lambda_a \left( |\Psi_{\pm 1}|^2 + |\Psi_{0}|^2 - |\Psi_{\mp 1}|^2 \right) \Psi_{\pm 1}
         \nonumber
         \\
         &
	+ \lambda_a \Psi_{0}^2\Psi_{\mp 1}^{*}, 
\nonumber
\\
 i \frac {\partial \Psi_{0}}{\partial t}=&
         \;	\mathcal{L}_{1D} \Psi_{0} + \varepsilon_{0} \Psi_{0}
         \nonumber         
         \\
         &
	+ \lambda_a \left( |\Psi_{-1}|^2 + |\Psi_{+1}|^2 \right) \Psi_{0}
	\\
	&
	+ 2 \lambda_a \Psi_{+1}\Psi_{-1}\Psi_{0}^{*},
\nonumber
\end{align}
where $\varepsilon_{j}$ is the Zeeman energy of the $j$th hyperfine state in transverse harmonic oscillator energy units $E_{j}= \varepsilon_{j} \hbar \omega_{\perp}$. The energy of an atom in a hyperfine state with electronic 
spin $1/2$ and the nuclear spin $I=3/2$ (for $^{87}$Rb), in an external magnetic field ${\bf B}$ is given by the Breit-Rabi equation~\cite{rabi,woodgate}:
\begin{align}
\label{eq:Zeeman}
E_{\pm 1}= & \; - \frac{\Delta_{\rm hf}}{8} \mp g_{\rm I} \mu_{\rm I} B 
- \frac{\Delta_{\rm hf}}{2} \sqrt{1\pm\alpha+\alpha^{2}},
\nonumber
\\
E_{0}= & \; - \frac{\Delta_{\rm hf}}{8} - \frac{\Delta_{\rm hf}}{2} \sqrt{1+\alpha^{2}}.
\end{align}
Here $\Delta_{\rm hf}$ is the hyperfine splitting at zero magnetic field and $\alpha \equiv (g_{\rm I} \mu_{\rm I} B + g_{\rm J} \mu_{\rm B} B)/\Delta_{\rm hf}$, 
where $\mu_{\rm I}$ is the nuclear magneton, $\mu_{\rm B}$ is the Bohr magneton, $g_{\rm I}$ and $g_{\rm J}$ are the gyro-magnetic ratios of electron and nucleus. The magnitude of the magnetic field is taken to be $B=1$ Gauss. For $^{87}$Rb~\cite{arimondo:review,salomon:values}:
\begin{align}
\Delta_{\rm hf} = & \; 6834.68261090434(3) \mbox{MHz},
\nonumber
\\
g_{\rm I} = & \; 0.9951414(10) \times 10^{-3},
\nonumber
\\
g_{\rm J} = & \; 2.00233113(20).
\nonumber
\end{align}
In order to remove fast phase rotation from the numerics, we rescale Eqs.~(\ref{eq:Zeeman}) to $E_{j}(B=0)=0$. This yields the following: $E_{\pm 1}= \; \frac{\Delta_{\rm hf}}{2} (1 - \sqrt{1\pm\alpha+\alpha^{2}}) \mp g_{\rm I} \mu_{\rm I} B $, and $E_{0}= \; \frac{\Delta_{\rm hf}}{2} (1 -  \sqrt{1+\alpha^{2}})$.

\begin{figure}
\includegraphics[width=10\textwidth,height=3.7cm,keepaspectratio]{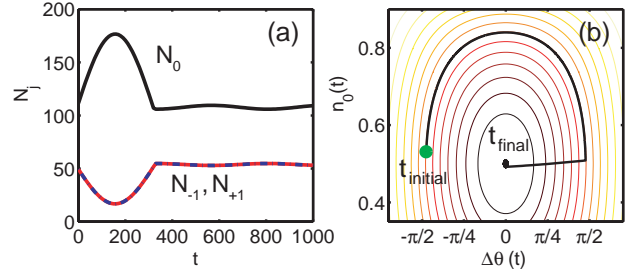}
\caption{\label{fig:halt}
Arrest of spin-mixing oscillations shown in \Fig~\ref{fig:noneq} due to quadratic Zeeman effect. The magnetic field of magnitude $B=1$ Gauss is applied over the period of $t_{B}=(\Delta \theta)/(\Delta E)\simeq 2$ ($1.4$~ms).
}
\end{figure}

\begin{figure}
\includegraphics[width=\columnwidth,keepaspectratio]{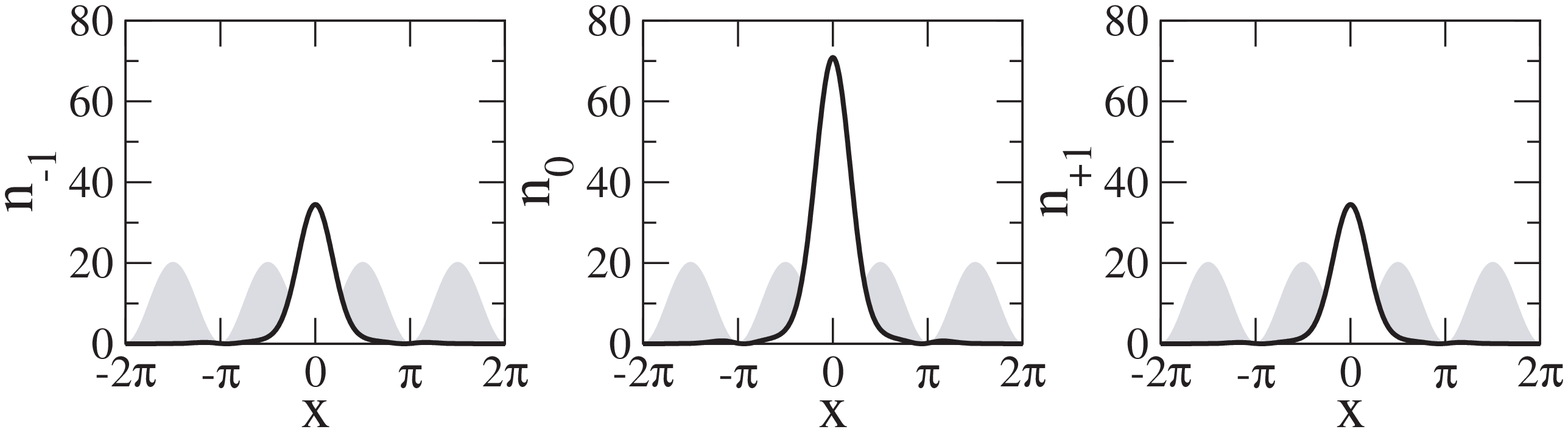}
\caption{\label{fig:haltdensity}
Density profiles of the hyperfine components of the vector gap soliton corresponding to $t_{\rm final}=1000$ in Fig.\ref{fig:halt}. Periodic lattice potential (shaded) has been increased by a factor of 5 for better visualization.
}
\end{figure}

If the Zeeman energy is much greater than other terms in equations~(\ref{eq:ZeemanGPEs}), the wave functions can be assumed to acquire phases $\theta_{j}(t) = -E_{j} t/\hbar$. The relative phase is then given by $\Delta \theta = 2 \theta_{0} - (\theta_{+1} +\theta_{-1})= -\Delta E t/\hbar$, where $\Delta E = 2 E_{0} - E_{-1} - E_{+1}$. From Eqs.~(\ref{eq:Zeeman}) it can then be seen that due to the {\em quadratic} Zeeman effect, the hyperfine states will acquire a change in the relative phase $\Delta \theta$. For a specific phase change, the duration of the magnetic field pulse $t_{B}$ is obtained as $t_{B}=\Delta \theta/ \Delta E$, which is around $t_{B}\simeq 2$ time units ($1.4$~ms) for the parameters chosen here, and is much smaller than the total duration of the evolution. The magnetic field pulse is initiated when the population $n_{0}$ approaches the value $1/2$. The resulting halt of the spin mixing oscillations is shown in~\Fig~\ref{fig:halt}. The soliton maintains its shape through the transition to the stationary state, with the final state shown in ~\Fig~\ref{fig:haltdensity}.

We point out that the advantage of studying the spin mixing dynamics with the gap solitons is twofold. First, due to the tight confinement in the lattice, solitons exhibit very regular and undisturbed oscillations. Secondly, the stability of the oscillations, together with the susceptibility to the magnetic field opens up the possibility for controlled spinor gap soliton manipulations.

\section{Conclusions}
We have modeled the nonlinear behaviour of spin-1 Bose-Einstein condensates with {\em ferromagnetic} ($^{87}\rm Rb$) and {\em polar} ($^{23}\rm Na$) types of spin-dependent interactions loaded into a one-dimensional optical lattice. By performing numerical analysis of the condensate evolution, we established that nonlinear periodic Bloch states formed by the condensate at the edge of the Brillouin zone are modulationally unstable, irrespective of the type of interactions. This instability underpins the existence of the multi-component, spatially localized states of the condensate inside the band gaps of the linear Bloch-wave spectrum. 

Furthermore, we have demonstrated that the localization of the spinor matter waves in an optical lattice takes the form of vector gap solitons and self-trapped waves. Both types of localized states may exist as in-phase and out-of-phase stationary spin configurations independent of whether the inter-atomic interactions are polar or ferromagnetic. The properties of these gap states are well described within the single-mode approximation frequently employed in the description of spinor systems.

We have also identified nonlinear stationary states that cannot be captured by the single mode approximation. These are ``hybrid'' gap states that are composed of a spatially localized soliton-like state in one of the hyperfine components and self-trapped states with a phase ``kink'' in the other components. We have examined stability properties of non-SMA states and found that {\it off-site} states exhibit Josephson-like oscillations between lattice wells, which eventually lead to a decay of the stationary state. Furthermore, we have shown that all the non-SMA states exhibit {\em self-magnetisation}, an intrinsic precession of the local spin.

Finally, we have shown that localized gap solitons prepared with a population or phase imbalance, i.e. in a non-stationary configuration, may display coherent undamped spin-mixing dynamics. Application of an external magnetic field can lead to the arrest of the inter-species population oscillations and rapid transfer to an equilibrium state. This opens up possibilities for controlled manipulation of the spinor condensate as a means of quantum state engineering, which is an essential element of quantum computing schemes.

\begin{acknowledgments}
\noindent BJDW thanks Sebastian W\"{u}ster for fruitful discussions and acknowledges help of Dr.~Andrew Truscott.
Authors express their gratitude to the Department of Physics, Faculty of Science of the Australian National University 
for kind hospitality. This research was supported by the Australian Research Council (ARC) and by a grant under the Supercomputer 
Time Allocation Scheme of the National Facility of the Australian Partnership for Advanced Computing. 
\end{acknowledgments}

\end{document}